# Surface optical phonon scattering in N-polar GaN quantum well channels

Uttam Singisetti*

*Electrical Engineering Department, University at Buffalo, Buffalo, NY, 14260, USA*

*Email: uttamsin@buffalo.edu, Ph: 716-645-1536, Fax: 716-645-3656*

N-polar GaN channel mobility is important for high frequency device applications. In this Letter, we report the theoretical calculations on the surface optical (SO) phonon scattering rate of two-dimensional electron gas (2-DEG) in N-polar GaN quantum well channels with high-k dielectrics. The effect of SO phonons on 2-DEG mobility was found to be small at >5 nm channel thickness. However, the SO mobility in 3 nm N-polar GaN channels with high-k dielectrics is low and limits the total mobility. The SO scattering for $SiN_x$ dielectric GaN was found to be negligible due to its high SO phonon energy.

The key advantage of the GaN devices is high two-dimensional electron gas (2-DEG) mobility in combination with wide bandgap which enables both high frequency and high breakdown operation [1]. Understanding the limiting mechanisms of the 2-DEG mobility [2-7] is critical as novel GaN device structures are explored. [8-11] In recent years there has been considerable interest and progress in the N-polar GaN field-effect transistors (FET) for ultra high frequency and high breakdown applications because of its inherent advantages of back barrier for electron confinement. [12-15] In the N-polar GaN devices, the quantum well channel thickness is critical to scaling of these devices to ultra-small gate lengths. [14, 16] In addition, the gate barrier also needs to be scaled in N-polar GaN devices by decreasing the thickness and also by incorporating high-k dielectrics. However, both the channel thickness scaling [17,18] and gate barrier scaling [4, 6] have been shown to affect the 2-DEG mobility which can impact the device performance. The channel thickness has been experimentally shown affect the channel velocity in devices. [16] Through device design and growth optimization, high conductivity 5 nm N-polar GaN channel and high mobilities ( 1100 cm$^2$/V.s) have been recently reported. [19-21] In order to continue further scaling of the high frequency performance in sub-20-nm gate length devices further channel thickness scaling (< 5 nm) with high 2-DEG mobility will be necessary. It is important to evaluate the limits of 2-DEG mobility arising from high-k dielectrics on these thin channels.

The deposition of high-k dielectrics on GaN results in remote interfacial charge scattering, [4, 6] remote roughness scattering. [2] Another important scattering mechanism originating from high-k dielectrics is the surface/interfacial polar optical phonon scattering, which has not been considered previously in N-polar GaN devices.

Surface optical (SO) phonon scattering, also referred to as remote interface phonon (RIP) and surface polar phonon (SPP), of the 2-DEG in Si MOSFET was reported by Hess [22] and Ferry [23]. The SO scattering was found to be a critical mobility limiting mechanism in Si, and III-V MOSFETs with high-k dielectrics. [24, 25] In recent years, SO phonon scattering has been identified as the limiting mechanism for room temperature mobility of graphene. [26-28] In this Letter, we report the impact of surface optical phonon (SOP) scattering on the mobility of N-polar channel devices. We also report on the dependence of the SO phonon scattering rate on the thickness of the N-polar GaN channel.

Surface or, interface optical phonon modes exist at the intersection of two materials if one of them is a polar material. [29, 30] These modes are confined to the interface and decay exponentially from it. In the case of high-k dielectric on semiconductor, the polar or ionic nature of the high-k leads surface/interface phonon modes. The interface modes arise due to the electrostatic boundary condition imposed on the phonon electric field. [30] Each transverse optical (TO) mode in the high-k dielectric gives rise to a SO mode at the interface. For high-k dielectrics on GaN there will be three interface modes, corresponding to the two polar phonons of the high-k and one polar optical phonon of GaN.

In order to calculate the SO phonon frequency we use electrostatic boundary condition for the metal-high-k-N-polar GaN system (Fig. 1). [24-26] To simplify calculations, we have ignored the role of the AlN bottom barrier in this analysis. For the metal dielectric function, it is considered to be a perfect metal without the plasmon contribution. Following the analysis in Ref. 26, applications of the electrostatic boundary condition gives rise to the secular equation for the interface phonon modes as

$$\varepsilon_{ox}(\omega)\cosh(qt_{ox}) + \varepsilon_{GaN}(\omega)\sinh(qt_{ox}) = 0, \qquad (1)$$

where, $\varepsilon_{ox}(\omega)$ and $\varepsilon_{GaN}(\omega)$ are the frequency dependent dielectric function of the high-k oxide and GaN respectively, which are given by

$$\varepsilon_{ox}(\omega) = \varepsilon_{ox}^{\infty} + \frac{f_1\omega_{\mathrm{TO1}}^2}{\omega_{\mathrm{TO1}}^2 - \omega^2} + \frac{f_2\omega_{\mathrm{TO2}}^2}{\omega_{\mathrm{TO2}}^2 - \omega^2}$$

and

$$\varepsilon_{GaN}(\omega) = \varepsilon_{GaN}^{\infty} + \frac{f_3\omega_{\mathrm{TO3}}^2}{\omega_{\mathrm{TO3}}^2 - \omega^2}$$

where the dielectric functions ($\varepsilon$) with superscripts of $\infty$ , $i$ and 0 correspond to high frequency, intermediate frequency, and DC dielectric constants. The coupling strengths are $f_1=\varepsilon^i_{ox}-\varepsilon^0_{ox}$, $f_2=\varepsilon^{\infty}_{ox}-\varepsilon^i_{ox}$, and $f_3=\varepsilon^{\infty}_{GaN} - \varepsilon^0_{GaN}$. The secular equation (Eq. 1) can be solved numerically to obtain the three solutions. The oxide thickness is taken to be 5 nm in this study. The high-k dielectric function parameters were taken from Ref. 25, the $SiN_x$ parameters[31, 32)] were taken from Ref. 32. The dispersion of the interface modes (SO1, SO2, and SO3) calculated from Eq (1) is shown in Fig. 2 for metal-$HfO_2$-GaN system. As seen in the figure, the SO modes are dispersion free except for at the low wave vectors. The small dispersion at low wave vectors is neglected in scattering rate calculations. The calculated SO phonon frequencies ($\omega_{SO}$) for all the dielectrics considered here are given in Table I.

For scattering rate calculations, the SO phonon-electron Fröhlic interaction Hamiltonian for a phonon with frequency $\omega_{SO}$ is given by [33)]

$$H_{int} = eF\sum_q\left[\frac{e^{-qz}}{\sqrt{q}}(e^{i\mathbf{q}\cdot\mathbf{r}}a_q^+ + e^{-i\mathbf{q}\cdot\mathbf{r}}a_q)\right],$$

where the interaction parameter F is [29)]

$$F = \left[ \left( \frac{1}{\varepsilon_{ox}^{\infty} + 1} - \frac{1}{\varepsilon_{ox}^{0} + 1} \right) \frac{\hbar \omega_{SO}}{2A\varepsilon_0} \right]^{1/2},$$

with $A$ being the normalization area, $\boldsymbol{q}$ is the two-dimensional phonon wave vector, $\boldsymbol{r}$ is the two-dimensional position vector, $\varepsilon_0$ is the free space permittivity, and $a^{+}_{q}(a_q)$ is the phonon annihilation (creation) operator. Now, the matrix element for single SO-phonon electron scattering process is calculated by $M(k,k') = <\psi_{kq} | H_{int} | \psi_{k'q}>$, where $\psi$ is the 2-DEG wavefunction, $\boldsymbol{k}$ and $\boldsymbol{k'}$ are the two-dimensional electron wave vector before and after scattering. The equation for the single phonon scattering rate is given by Fermi's golden rule

$$S(\boldsymbol{k},\boldsymbol{k'}) = \frac{2\pi}{\hbar} | M(\boldsymbol{k},\boldsymbol{k'}) |^2 \, \delta(\boldsymbol{k},\boldsymbol{k} \pm \boldsymbol{q}) \delta(E' - E \mp \hbar\omega_{SO}), \qquad (2)$$

where the two $\delta$-functions are for the conservation of the momentum and energy during the scattering process, $E$ and $E'$ are the initial and final electron energy. The product of the two $\delta$-functions can be converted to a single $\delta$-function [34)]

$$\delta(\boldsymbol{k},\boldsymbol{k} \pm \boldsymbol{q}) \delta(E' - E \mp \hbar \boldsymbol{\omega_{SO}}) \rightarrow \frac{1}{\hbar v q} \delta(\pm \cos\theta + \frac{\hbar q}{2p} \mp \frac{\omega_{SO}}{v q}),$$

where $\theta$ is the angle between $\boldsymbol{k'}$ and $\boldsymbol{q}$, $v$, and $p$ are the electron velocity and momentum respectively. The relationship between the electron wavevectors and phonon wave vector is shown in Fig 2 inset. The total SO phonon scattering momentum relaxation rate is evaluated by summing over all the finals states $\boldsymbol{k'}$ which satisfy the momentum and energy conservation, or equivalently summing over all the $\boldsymbol{q}$ states that satisfy the momentum and energy conservation.[26, 34)]

$$\frac{1}{\tau_m}=\frac{2\pi}{\hbar}\sum_q\left|\frac{M(\boldsymbol{k},\boldsymbol{k'})}{\varepsilon_{2D}(q)}\right|^2(1-\frac{k'}{k}\cos\alpha)(N_q+\frac{1}{2}\mp\frac{1}{2})\frac{1}{\hbar\upsilon q}\delta(\pm\cos\theta+\frac{\hbar q}{2p}\mp\frac{\omega_{SO}}{\upsilon q}), \qquad (3)$$

where $\varepsilon_{2D}$ is the Thomas-Fermi screening factor for the 2-DEG with a finite thickness [17], and $N_q$ is the Bose-Einstein phonon occupation number at temperature $T$. The term in the first parenthesis in Eq. 3 is the momentum randomization factor. The upper sign is for phonon absorption and the lower sign is for phonon emission. The summation over $q$ can be converted to an integral ($\sum_q(\cdot)\rightarrow\frac{1}{4\pi^2}\int qdq\int d\theta$) giving the equation for the scattering rate as

$$\frac{1}{\tau_m}=\frac{2\pi}{\hbar}\frac{1}{4\pi^2}\int\left|\frac{M(\boldsymbol{k},\boldsymbol{k'})}{\varepsilon_{2D}(q)}\right|^2(1-\frac{k'}{k}\cos\alpha)(N_q+\frac{1}{2}\mp\frac{1}{2})\frac{1}{\hbar\upsilon q}qdq\int\delta(\pm\cos\theta+\frac{\hbar q}{2p}\mp\frac{\omega_{SO}}{\upsilon q})d\theta. \qquad (4)$$

From Fig. 3 the momentum randomization factor in the parenthesis can be written in terms of $q$ and angle θ. [34] The $\delta$-function integral can be converted to an integral over $q$, [34] which gives the final expression for the SO phonon scattering rate as

$$\frac{1}{\tau_m}=\frac{1}{2\pi\hbar\upsilon p}\int_{q_{min}}^{q_{min}}\left|\frac{M(\boldsymbol{k},\boldsymbol{k'})}{\varepsilon_{2D}(q)}\right|^2 f(q)(N_q+\frac{1}{2}\mp\frac{1}{2})qdq,$$

$$\text{where } f(q)=\frac{\left(\frac{\hbar q}{2p}\mp\frac{\omega_{SO}}{vq}\right)}{\left(\sqrt{1-\left(\frac{\hbar q}{2p}\mp\frac{\omega_{SO}}{vq}\right)^2}\right)}$$

The limits on the phonon wave vectors $q_{min}$ and $q_{max}$ are obtained from the momentum and energy conservation equations and are given by [34]

$$q_{min} = \frac{p}{\hbar}\left(1 \pm \sqrt{1 \pm \frac{\hbar\omega_{SO}}{E}}\right), and\ q_{max} = \frac{p}{\hbar}\left(\mp 1 \pm \sqrt{1 \pm \frac{\hbar\omega_{SO}}{E}}\right),$$

where $E$ is the electron energy. For phonon emission process $E>\hbar\omega_{SO}$, this ensures the phonon wave vector limits are real. Eq. 4 can now be evaluated numerically to get SO phonon scattering rate.

In order to calculate the scattering rate in N-polar quantum well channels, we use the variational wavefunction for the 2-DEG [17, 35)] with the coordinate system shown in Fig. 1

$$\psi(z) = N(\beta)\sin\left(\frac{\pi z}{L}\right)\exp\left(-\beta\left(1-\frac{z}{L}\right)\right),\ \ 0 < z < L$$

Where $L$ is the channel thickness, $N^2(\beta) = 4\beta(\beta^2+\pi^2)\exp(2\beta)/L\pi^2(\exp(2\beta)-1)$ is the normalization constant, and $\beta_{min} = ((3/4)\pi^2 eF_{avg}L/(\hbar^2\pi^2/2m^*L^2))^{1/3}$ is the channel electric field dependent ($F_{avg}$) variational parameter. The electric field in the N-polar GaN channel device with a 2-DEG density of $n_s$, is taken to be $F_{avg} = e(N_{2D} - n_s/2)/\varepsilon^0_{GaN}$, where $N_{2D}$ is the total two-dimensional modulation doping density. $N_{2D}$ is taken to be $1\times10^{13}$ cm$^{-2}$ for the scattering calculations reported here. With the wave function determined, the scattering matrix element can be evaluated from the Eq. 4. The electron mobility limited by SO phonon is derived by $e\tau_m/m^*$, where $m^*$=0.2 is the GaN electron effective mass.

Fig.3 (a) shows the momentum relaxation rates for different dielectrics on 5 nm N-polar GaN channel. There is sharp increase in the SO phonon scattering rate when phonon emission process becomes feasible at electron energies greater than the SO energy. The phonon scattering rate for $Al_2O_3$ is lower than that of $HfO_2$ and $ZrO_2$ layers till the electron energy reaches the $Al_2O_3$ SO1 phonon energy (54 meV). At $E$>54 meV, the $Al_2O_3$ SO1 phonon emission process becomes feasible and the scattering rate of $Al_2O_3$, $HfO_2$ and $ZrO_2$ become comparable. The figure also shows that the SO phonon scattering rate is an order lower for $SiN_x$ dielectric layer which has a very high SO phonon energy of 148 meV. The high SO phonon energy suppresses of the phonon emission process in $SiN_x$ dielectric/GaN channel layers. In addition, it also reduces the phonon occupancy number ($N_q$), resulting in a low phonon scattering rate.

Fig. 3 (b) shows the SO phonon limited mobility for the 5 nm GaN channel as a function of 2-DEG concentration. At 2-DEG densities < $4.5\times10^{12}$, the SO phonon limited mobility is highest for $Al_2O_3$, above which the mobility drops to a value comparable to $HfO_2$ and $ZrO_2$ dielctrics. At 2-DEG densities > $7\times10^{12}$, when both SO1 and SO2 phonon emission process is allowed for $Al_2O_3$ it has lower mobility than $HfO_2$ and $ZrO_2$. The SO limited mobility depends on the 2-DEG density and the high-k dielectrics with different materials giving the highest mobility at different 2-DEG densities. The $SiN_x$ phonon limited mobility is > $10^5$ $cm^2/V.s$ for all the 2-DEG densities due to the lower scattering rate.

The total mobility of the 2-DEG will be significantly affected if the SO phonon mobility becomes comparable to the room temperature GaN bulk phonon limited mobility which is of the order of 2000 $cm^2/V.s$ [2]. The lowest SO limited phonon for the 5 nm channel is 4000 $cm^2/V.s$ for $Al_2O_3$ dielectrics at 2-DEG density of $6.5\times10^{12}$ $cm^{-2}$ which would drop that total 2-DEG mobility to 1350 $cm^2/V.s$. At other 2-DEG densities a higher SO phonon limited mobility is expected. The best reported mobility for 5 nm N-polar GaN channel is 1100 $cm^2/V.s$, [19-21)] which would not be significantly degraded by the SO scattering rates calculated here. It could be concluded that the SO limited phonon scattering is not a significant mobility limiting mechanism at 5 nm GaN channel thickness. At 8 nm GaN channel thickness the SO phonon limited mobility was found to be > $4\times10^4$ $cm^2$/V.s for all the dielectrics because of the negligible overlap of 2-DEG with the SO phonons.

Next, we explore the SO phonon limited mobility for 3 nm GaN channel. Fig. 4 (a) shows the calculated SO limited mobility for 3 GaN channel devices with 2-DEG density as a function of 2-DEG density. It is observed that the SO limited mobility is < 2000 $cm^2$/V.s for $Al_2O_3$, $HfO_2$ and $ZrO_2$ dielectrics for 2-DEG densities > $4.5\times10^{12}$ $cm^{-2}$. The lower mobility in 3-nm GaN channels is due to the increased scattering rate resulting from the increased overlap between the 2-DEG wave function and the SO phonons. This will severely degrade the 2-DEG mobility in 3 nm channel thickness and will be a limiting factor for high frequency device applications.

In Fig. 4(b), the calculated SO phonon limited mobility for $SiN_x$ is shown to be > 20000 $cm^{-2}$/V.s for all 2-DEG densities. The higher mobility using $SiN_x$ dielectric is due to the lower occupancy of the $SiN_x$ SO phonons and also due to the suppression of the phonon emission process. However, because of the lower dielectric constant of $SiN_x$, it reduces the device EOT, which is a critical parameter for scaled devices. A composite dielectric layer such as $SiN_x/Al_2O_3$, $SiN_x/HfO_2$ could be used to reduce the SO phonon scattering in these ultra-thin channels. In addition, a thin AlN layer could also be used as an interfacial layer because it has a large SO phonon frequency.

The SO limited phonon mobility is a difficult challenge at 3 nm GaN channel devices with high-k. The model presented here could be used to design the high-k/channel structures that can be used for extremely scaled ( < 5nm) N-polar GaN channels. Comprehensive experimental evaluation of the 2-DEG mobility of thin GaN channel devices can be carried out using single and multi-layer dielectrics to verify the predictions in these calculations.

In summary, we have calculated the SO phonon scattering rates of $Al_2O_3$, $ZrO_2$, $HfO_2$ and $SiN_x$ dielectrics on GaN quantum well channels. The SO phonon limited mobility at channel thickness 8 nm was order of magnitude larger than GaN bulk phonon limited mobility. The SO mobility at 5 nm thickness is found to > 4X higher than the experimentally reported mobilities [19, 20]. However, the SO limited mobility becomes a limiting factor at 3 nm GaN channel thickness and will severely degrade the 2-DEG mobility for all the considered high-k dielectrics except for $SiN_x$. By careful design of composite gate stack the incorporate interfacial SiN or AlN layer with high SO phonon energy can be used to improve the mobility in thin (< 5 nm) channels.

**Acknowledgement**: The author thanks Dr. Man Hoi Wong for carefully reading the manuscript and helpful suggestions.

**References**:

1. U. K. Mishra, P. Parikh and Y.-F. Wu: Proc. IEEE **90** (2002) 1022.
2. Y. Cao, H. Xing and D. Jena: Appl. Phys. Lett. **97** (2010) 222116.
3. D. Jena, A. C. Gossard and U. K. Mishra: Appl. Phys. Lett. **76** (2000) 1707.
4. T.-H. Hung, M. Esposto and S. Rajan: Appl. Phys. Lett. **99** (2011) 162104.
5. D. Ji, Y. Lu, B. Liu, G. Jin, G. Liu, Q. Zhu and Z. Wang: J. Appl. Phys. **112** (2012) 024515.
6. D. Ji, B. Liu, Y. Lu, G. Liu, Q. Zhu and Z. Wang: Appl. Phys. Lett. **100** (2012) 132105.
7. D. F. Brown, S. Rajan, S. Keller, Y.-H. Hsieh, S. P. DenBaars and U. K. Mishra: Appl. Phys. Lett. **93** (2008) 042104.
8. D. S. Lee, L. Bin, M. Azize, G. Xiang, G. Shiping, D. Kopp, P. Fay and T. Palacios: IEDM Tech. Dig. (2011), p. 19.12.11.
9. B. Lu, E. Matioli and T. Palacios: IEEE Electron Device Lett. **33** (2012) 360.
10. K. Shinohara, D. Regan, A. Corrion, D. Brown, I. Alvarado-Rodoriguez, M. Cunningham, C. Butler, A. Schmitz, S. Kim, B. Holden, D. Chang, V. Lee and P. M. Asbeck: IEEE Compound Semiconductor Integrated Circuit Symposium (CSICS) (2012), p. 1.
11. Y. Yue, Z. Hu, J. Guo, B. Sensale-Rodriguez, G. Li, R. Wang, F. Faria, T. Fang, B. Song, X. Gao, S. Guo, T. Kosel, G. Snider, P. Fay, D. Jena and H. Xing: IEEE Electron Device Lett. **33** (2012) 988.
12. S. Rajan, A. Chini, M. H. Wong, J. S. Speck and U. K. Mishra: J. Appl. Phys. **102** (2007), 044501.
13. Nidhi, S. Dasgupta, D. F. Brown, S. Keller, J. S. Speck and U. K. Mishra: IEDM Tech. Dig., (2009), p. 1-3.
14. U. Singisetti, M. H. Wong, J. S. Speck and U. K. Mishra: IEEE Electron Device Lett. **33** (2012) 26.
15. D. Denninghoff, J. Lu, M. Laurent, E. Ahmadi, S. Keller and U. K. Mishra: IEEE Device Research Conference (DRC) (2012), p. 151.
16. Nidhi, S. Dasgupta, J. Lu, J. S. Speck and U. K. Mishra: IEEE Electron Device Lett. **33** (2012) 961.
17. U. Singisetti, M. H. Wong and U. K. Mishra: Appl. Phys. Lett. **101** (2012) 012101.
18. Nidhi, O. Bierwagen, S. Dasgupta, D. F. Brown, S. Keller, J. S. Speck and U. K. Mishra: Electronic Materials Conferece (EMC) (2010).
19. D. Denninghoff, J. Lu, E. Ahmadi, S. Keller and U. K. Mishra: 39th International Symposium on Compound Semiconductors, (2012).
20. J. Lu, M. Laurent, R. Chung, S. Lal, A. Sztein, S. Keller, S. P. DenBaars and U. K. Mishra: Electronic Materials Conferece (EMC) (2012).
21. J. Lu, D. Denninghoff, S. Keller, S. P. DenBaars and U. K. Mishra: International Workshop on Nitride Semiconductors (2012).
22. K. Hess and P. Vogl: Solid State Communications **30** (1979) 797.
23. B. T. Moore and D. K. Ferry: Journal of Applied Physics **51** (1980) 2603.
24. T. P. O'Regan, M. V. Fischetti, B. Soree, S. Jin, W. Magnus and M. Meuris: J. Appl. Phys. **108** (2010) 103705.
25. M. V. Fischetti, D. A. Neumayer and E. A. Cartier: J. Appl. Phys. **90** (2001) 4587.
26. A. Konar, T. Fang and D. Jena: Phys. Rev. B **82** (2010) 115452.
27. Z.-Y. Ong and M. V. Fischetti: Phys. Rev. B **86** (2012) 165422.

28. S. Fratini and F. Guinea: Phys. Rev. B **77** (2008) 195415.
29. S. Q. Wang and G. D. Mahan: Phys. Rev. B **6** (1972) 4517.
30. P. Y. Yu and M. Cardona: *Fundamentals of Semiconductors: Physics and Materials Properties* (Springer Berlin Heidelberg, 2010), p. 502.
31. D. V. Tsu, G. Lucovsky and M. J. Mantini: Phys. Rev. B **33** (1986) 7069.
32. O. Debieu, R. P. Nalini, J. Cardin, X. Portier, J. Perriere and F. Gourbilleau: Nanoscale Research Lett. **8** (2013) 31.
33. T. Fang, A. Konar, H. Xing and D. Jena: Phys. Rev. B **84** (2011) 125450.
34. M. Lundstrom: *Fundamentals of Carrier Transport* (Cambridge University Press, 2000), p. 54-.
35. D. Ahn and S. L. Chuang: Appl. Phys. Lett. **49** (1986) 1450.

**Figure Captions:**

Fig. 1 : (a) Schematic cross-section of the N-polar GaN device. (b) Schematic conduction band profile of the device. The figure shows the z-co-ordinate used for the calculations.

Fig. 2: Calculated dispersion of SO phonon modes for metal-$HfO_2$-GaN system. Bottom two modes arise from the $HfO_2$ optical modes and the top mode originates from GaN optical mode. (Inset) Momentum conservation during the SO phonon absorption process.

Fig. 3: (a) Calculated SO phonon momentum relaxation rate for various dielectrics for a 5 nm GaN channel. (b) Calculated SO phonon limited mobility for 5 nm GaN channel device for $Al_2O_3$, $HfO_2$ and $ZrO_2$ high-k dielectrics.

Fig. 4: (a) Calculated SO phonon limited mobility for 3 nm GaN channel device for $Al_2O_3$, $HfO_2$ and $ZrO_2$ high-k dielectrics. (b) Calculated SO phonon limited mobility for 3 nm GaN channel device for $SiN_x$ dielectric.

**Table Captions:**

TABLE I. Calculated SO phonon frequencies for various dielectrics with $t_{ox}$=5nm. The dielectric function paramters for $Al_2O_3$, $HfO_2$ and $ZrO_2$ were taken from Ref. 25. For $SiN_x$, single SO mode is obtained because of the single TO phonon in the bulk $SiN_x$. The dielectric function paramters for $SiN_x$ are $\varepsilon^{\infty}_{SiN}$=7.5, $\varepsilon^{0}_{SiN}$=4.17, and $\omega_{TO1}$=103 meV. [32)]

| | $Al_2O_3$ | $HfO_2$ | $ZrO_2$ | $SiN_x$ |
|---|---|---|---|---|
| $\omega_{SO1}$ (meV) | 54.0 | 19.30 | 25.10 | 148.0 |
| $\omega_{SO2}$ (meV) | 80.0 | 51.50 | 63.30 | - |

Singisetti , Fig. 1:

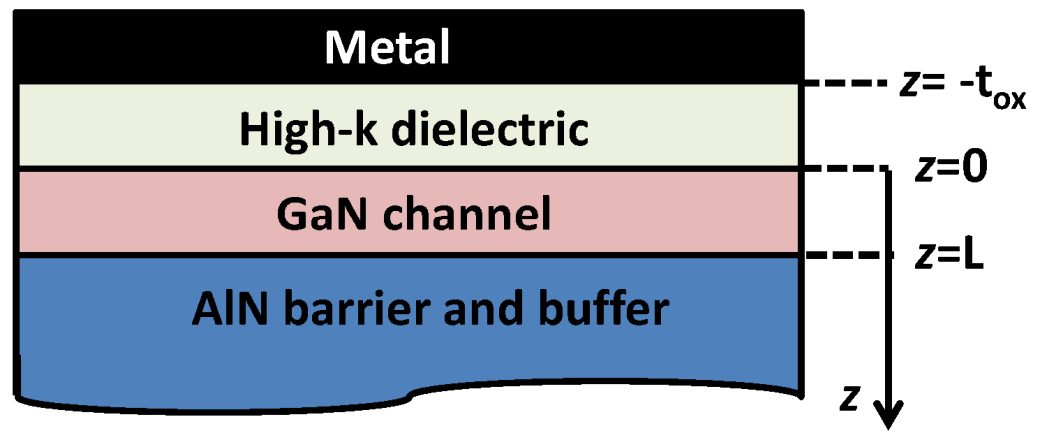

(a)

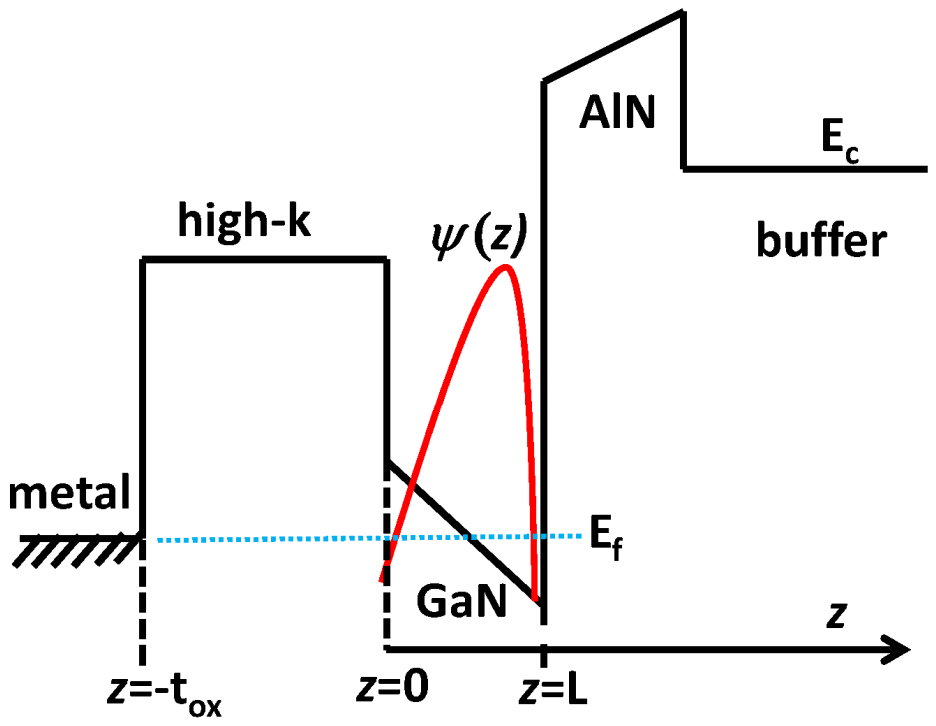

(b)

Singisetti , Fig. 2:

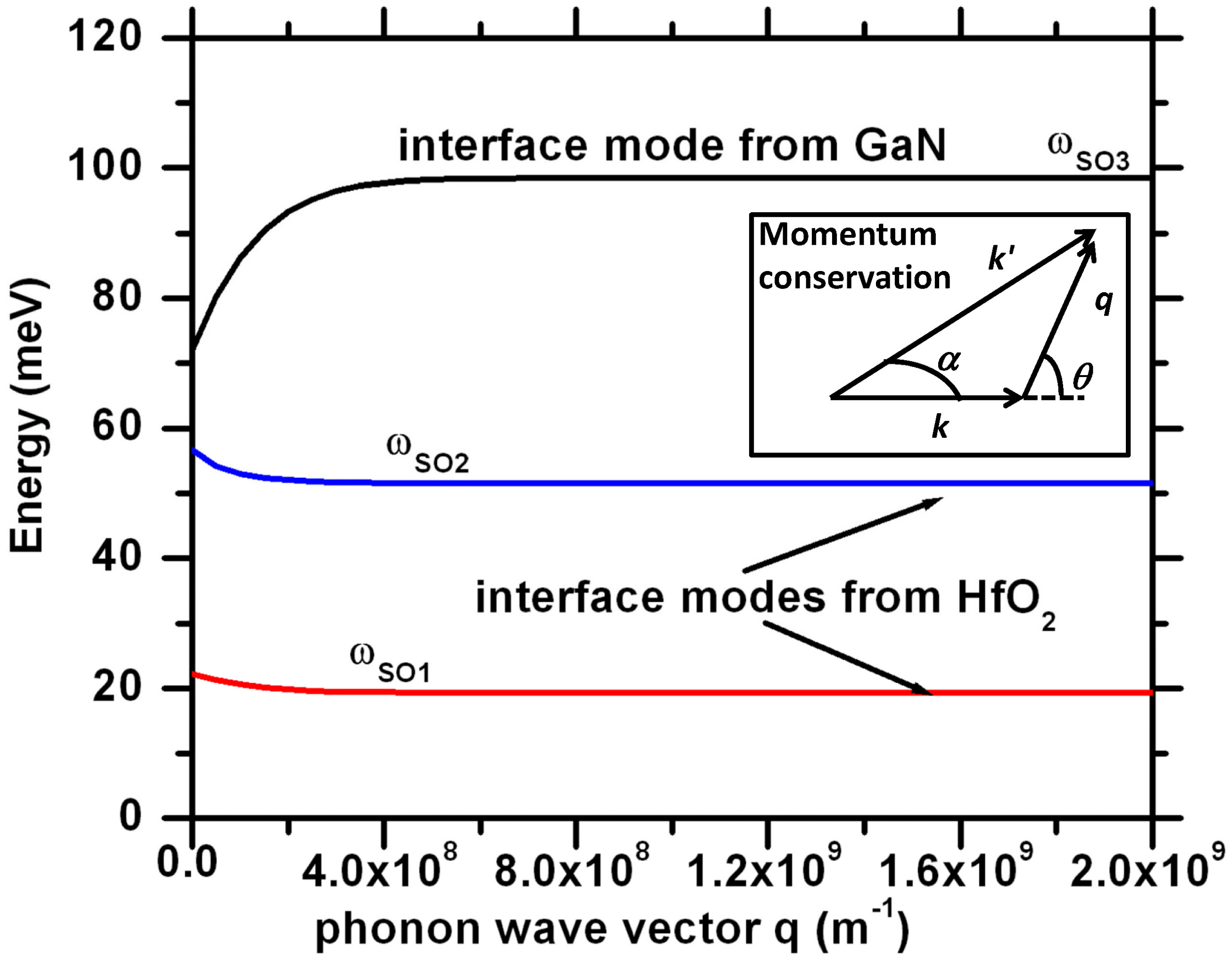

Singisetti , Fig. 3:

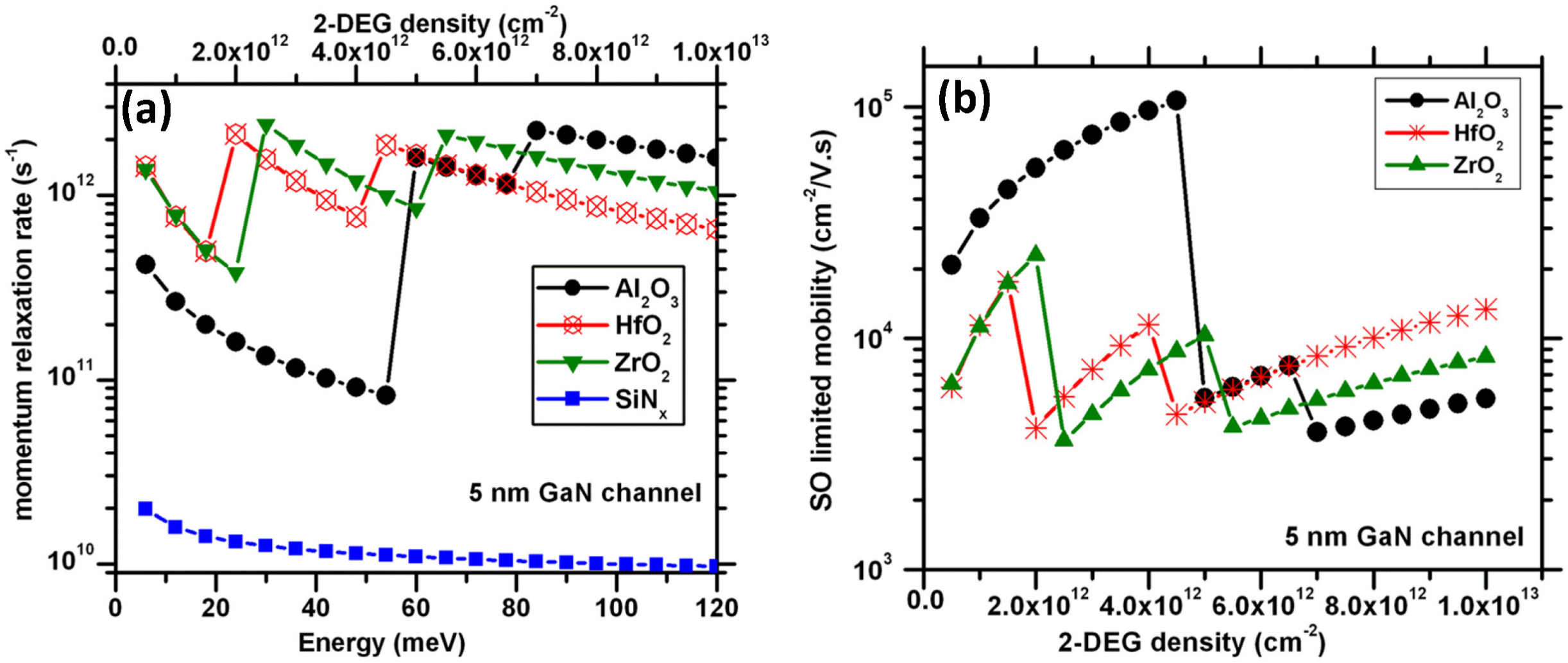

Singisetti , Fig. 4:

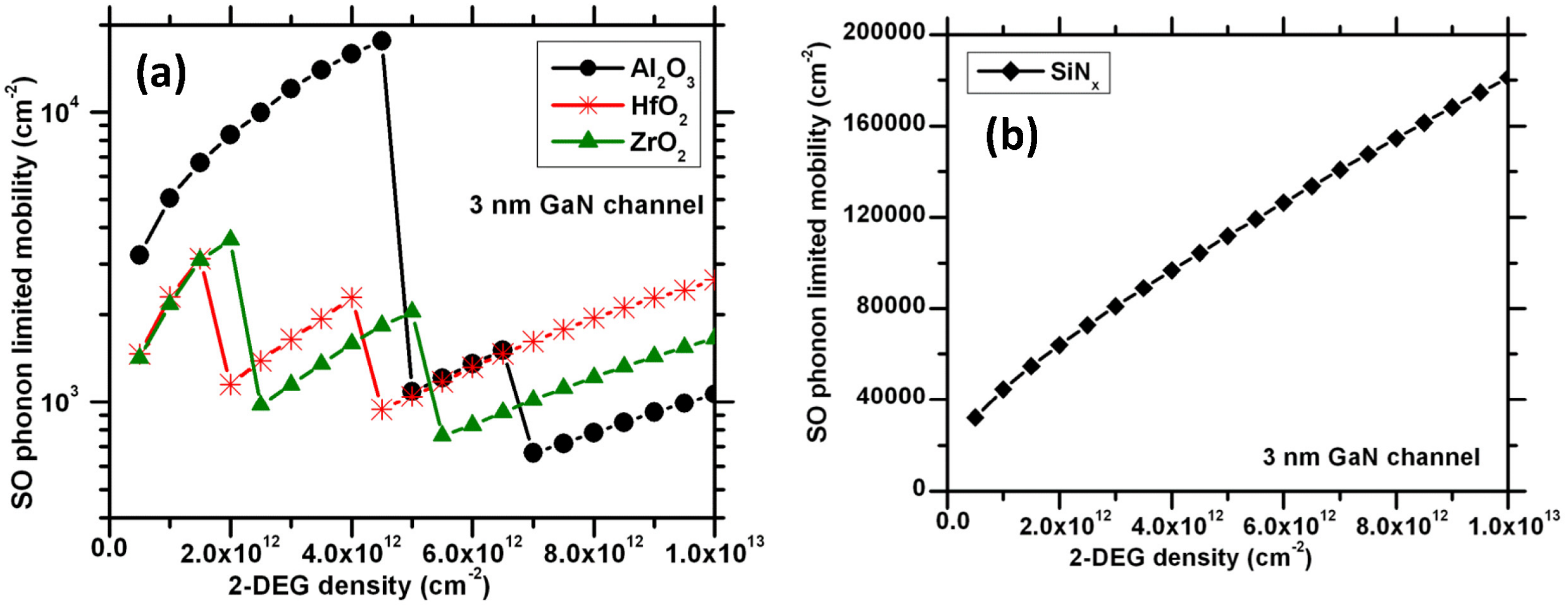